\documentclass[9pt,article,twoside]{rmaa-rho-class/rmaa-rho}
\RMxAAtemplatetype{\RMxAA} 

\newcommand\rmaatex{RMxAA~\LaTeX}

\vol{62-2}
\pages{XX-XX}
\thisyear{2026}
\doi{\href{https://www.astroscu.unam.mx/rmaa/RMxAA..XX-X}{https://www.astroscu.unam.mx/rmaa/RMxAA..XX-X}}

\title{Optical Design of OPTICAM-ARG: A Three-Channel High-Time-Resolution Camera for the Jorge Sahade Telescope}
 
\author[1]{M.~R.~ Nájera \orcidlink{0000-0003-3283-0407}}
\author[1]{J.~Herrera~Vázquez \orcidlink{0000-0002-1654-3283}}
\author[1]{E.~Luna \orcidlink{0000-0002-4056-1887}}
\author[1]{A.~Castro \orcidlink{0000-0002-7832-5337}}
\author[2]{D.~Altamirano \orcidlink{0000-0002-3422-0074}}
\author[1]{R.~Michel \orcidlink{0000-0003-1263-808X}}
\author[3,4]{S.~A.~Cellone \orcidlink{0000-0002-3866-2726}}
\author[1]{E. Sohn}

\affil[1]{Instituto de Astronomía, Universidad Nacional Autónoma de México, AP 106,  Ensenada 22800, BC, México}
\affil[2]{School of Physics and Astronomy, University of Southampton, Southampton, SO17 1BJ, UK}
\affil[3]{Complejo Astronómico El Leoncito (CASLEO), CONICET-UNLP-UNC-UNSJ, Av. España 1512 (Sur), J5402DSP, San Juan, Argentina}
\affil[4]{Facultad de Ciencias Astronómicas y Geofísicas, Universidad Nacional de La Plata, Paseo del Bosque, B1900FWA, La Plata, Argentina}

\leadauthor{Najera et al.}
\smalltitle{LaTex Macro RMxAA}

\corres{Morgan Rhaí Najera Roa}
\email{mnajera@astro.unam.mx}

\received{February 21, 2026}
\accepted{April 7, 2026}

\license{Texto de la licencia aquí}
\usepackage{float}

\setbool{rho-abstract}{true} 
\setbool{rho-resumen}{true} 

\begin{abstract}

    We present the optical design of \textit{OPTICAM-ARG}, a multi-channel instrument for the simultaneous acquisition of images in three spectral bands at the Cassegrain focus of an $f/8.5$ telescope, covering the $0.35$ to $1.00\,\mu$m wavelength range. The converging beam delivered by the telescope is spectrally separated by two dichroics into three channels, blue, green, and red, each incorporating a dedicated three-lens focal reducer, an interchangeable SDSS filter stage, and an sCMOS detector. The focal reducers establish an effective focal length of $\sim9.1$\,m, a uniform plate scale of $22.6\,\arcsec\,\mathrm{mm}^{-1}$, and a field of view of $8.4\,\arcmin\,\times\,8.4\,\arcmin$ per channel, consistent with the typical seeing conditions at the site. Operation of the dichroics in a converging beam introduces off-axis aberrations, which are mitigated through wedge angles applied to their second surface and optimized as part of the global design. Optical performance is assessed through exact ray tracing using RMS spot radii and encircled energy metrics, with EE50 values further expressed in terms of an equivalent FWHM to enable direct comparison with atmospheric seeing and to evaluate sensitivity to manufacturing tolerances.

\end{abstract}

\keywords{instrumentation: photometers --- techniques: photometric --- telescopes --- methods: observational}

\begin{resumen}

    Presentamos el diseño óptico de \textit{OPTICAM-ARG}, un instrumento multi-canal para la adquisición simultánea de imágenes en tres bandas espectrales en el foco Cassegrain de un telescopio $f/8.5$, que cubre el intervalo espectral de $0.35$ a $1.00\,\mu$m. El haz convergente entregado por el telescopio se separa espectralmente mediante dos dicroicos en tres canales, azul, verde y rojo, cada uno de los cuales incorpora un reductor focal dedicado de tres lentes, una etapa de filtros SDSS intercambiables y un detector sCMOS. Los reductores focales establecen una longitud focal efectiva de $\sim9.1$\,m, una escala de placa uniforme de $22.6\,\arcsec\,\mathrm{mm}^{-1}$ y un campo de visión de $8.4\,\arcmin\,\times\,8.4\,\arcmin$ por canal, en concordancia con las condiciones típicas de \textit{seeing} del sitio. La operación de los dicroicos en un haz convergente introduce aberraciones fuera de eje, las cuales se mitigan mediante ángulos de acuñamiento aplicados a su segunda superficie y optimizados como parte del diseño global. El desempeño óptico se evalúa mediante trazado exacto de rayos utilizando radios RMS del diagrama de manchas y métricas de energía encerrada, expresándose adicionalmente los valores de EE50 en términos de un FWHM equivalente para permitir una comparación directa con el \textit{seeing} atmosférico y evaluar la sensibilidad a tolerancias de fabricación.

\end{resumen}

\begin{document}

\maketitle
\pagestyle{fancy}\thispagestyle{firststyle}


\section{Introduction}

High-cadence astronomical observations across multiple spectral bands have become an essential tool for the study of transient and variable phenomena, including eclipsing binaries, interacting compact systems, and cataclysmic stars \citep{dhillon2021hipercam, 2007MNRAS.378..825D, butler2012first, castro-segura-2025}. In such programs, the characteristics of the instrumentation are not merely technical constraints but fundamental determinants of scientific feasibility, as they directly impact photometric precision, temporal continuity, and spectral simultaneity.

Despite their scientific importance, instruments capable of performing simultaneous multi-band, high-cadence photometry remain relatively scarce, particularly in the southern hemisphere. This limitation becomes increasingly relevant in the era of large-scale surveys such as the Vera C. Rubin Observatory Legacy Survey of Space and Time (LSST), which provides deep and wide-field coverage but is not optimized for targeted, continuous, and flexible monitoring of specific transient sources. In this context, small- and medium-aperture telescopes equipped with dedicated multichannel high-cadence instrumentation play a complementary and strategic role, offering temporal availability and operational flexibility that enable dedicated high-cadence campaigns \citep{miller2025lasillaschmidtsouthern}.

However, many of these facilities were conceived under historical optical design criteria, typically characterized by relatively slow focal ratios and limited fields of view. Such configurations are not always compatible with modern large-format detectors or with the performance requirements imposed by simultaneous multichannel operation. Adapting these telescopes to contemporary scientific demands, therefore, requires careful optical redesign rather than simple instrumental upgrades. From an optical design perspective, image formation is central to this adaptation process. In many astronomical instruments, image quality is understood in the classical geometric sense as the spatial concentration of energy and the faithful reproduction of the spatial structure of the object onto the detector plane \citep{sasian2019introduction}. Achieving this performance across multiple spectral channels introduces additional complexity, particularly when focal reduction and beam splitting must be implemented within constrained mechanical volumes.

A widely adopted strategy for extending the capabilities of existing telescopes is the incorporation of focal reducers, which increase the effective field of view and improve system efficiency without altering the primary optical configuration \citep{wilson2004reflecting, rutten1988telescope}. In multi-band instruments, this approach is complemented by dichroic elements that spectrally separate the converging beam. The insertion of dichroics into such beams, however, breaks rotational symmetry and introduces off-axis aberrations that must be carefully controlled, as discussed in Section\,\ref{subsec:OpticalConfiguration}. The design of multichannel astronomical cameras thus constitutes a coupled optical and methodological problem. Conventional purely numerical optimization strategies, frequently based on generic merit functions, may converge toward solutions that lack physical interpretability and exhibit strong dependence on arbitrarily chosen initial configurations \citep{malacara2003handbook, hoschel2019genetic}. This dependence can obscure the relationship among the first-order power distribution, third-order aberration control, and the final geometric image quality.

Multi-channel, high-time-resolution photometric systems like \textit{OPTICAM} \citep{castro+19} and \textit{HiPERCAM} \citep{dhillon2021hipercam} provide a powerful capability for time-domain astronomy, as they enable strictly simultaneous measurements in multi-wavelength bands, minimizing uncertainties associated with stochastic variability and atmospheric fluctuations. This simultaneous multi-band approach is particularly valuable for characterizing rapid variability, detecting periodic and quasi-periodic oscillations, measuring inter-band time lags, and tracking color-dependent flux changes in compact binaries, pulsating stars, and other transient phenomena. Such capabilities are essential for the study of fast optical variability and short-timescale behaviour reported in recent time-domain investigations \citep[e.g.][]{Irving+25, Castro-Segura+2025, Veresvarska+25, Shahbaz+15}.

In this work, we present the baseline optical design of \textit{OPTICAM-ARG}, a multi-band astronomical camera optimized for implementation at the Jorge Sahade\footnote{\url{https://casleo.conicet.gov.ar/js/}} (JS) Telescope of the El Leoncito Astronomical Complex (CASLEO), intended for high-cadence time-domain astronomy studies. The optical configuration is driven by the geometric and mechanical constraints imposed by the telescope, together with the image quality and spectral coverage requirements of high-cadence multi-band observations. Developed through a collaboration between the University of Southampton, the Universidad Nacional Autónoma de México, and CASLEO, the instrument preserves the multichannel philosophy of \textit{OPTICAM}, enabling simultaneous acquisition in three spectral bands while adapting the design to the optical and mechanical conditions of the JS Telescope.

The design is developed through a structured methodology that preserves physical coherence from the construction of the starting-point configuration through numerical refinement, explicitly controlling aberration balance and convergence behavior. The workflow integrates an open-source computational framework with complementary commercial verification tools, prioritizing reproducibility and interpretability throughout the design process. The paper is organized as follows. Section\,\ref{sec:Metodologia} describes the adopted optical design methodology. Section\,\ref{sec:Resultados} presents the nominal performance of the system, the effect of dichroic wedging, and the tolerance analysis. Finally, Sections\,\ref{sec:Discusion} and\,\ref{sec:Conclusiones} discuss the implications of the design and summarize the main conclusions.

\section{Optical configuration of \textit{OPTICAM-ARG}}\label{subsec:OpticalConfiguration}

    \textit{OPTICAM-ARG} is being developed for operation at the Cassegrain focus of the 2.15\,m Jorge Sahade telescope at CASLEO. The available mechanical envelope at the focal station defines the allowable instrument volume and imposes strict constraints on optical layout and component placement. The system employs two dichroic beam splitters in a converging beam to separate the incoming light into three simultaneous spectral channels (blue, green, and red). 
    
    At the beginning of the optical design, an alternative architecture based on a collimator with dichroics was considered, as it simplifies spectral splitting by avoiding wedge compensation. However, the adopted design places the dichroics in a converging beam followed by focal reducers, enabling a more compact layout by reducing the number of elements and overall system length, albeit at the expense of reduced geometrical flexibility compared to the collimated approach. Additionally, minimizing the number of optical surfaces improves throughput. Although this configuration introduces additional aberrations and requires wedged dichroics, these effects are explicitly accounted for and corrected during the optimization process. Therefore, the selected configuration represents a balanced trade-off between optical performance, compactness, and efficiency.

    Each channel includes a dedicated focal reducer, an interchangeable SDSS filter, and a back-illuminated scientific Complementary Metal-Oxide-Semiconductor (sCMOS) detector optimized for high frame-rate operation. The focal reduction factor and resulting plate scale were selected to provide appropriate sampling under typical site seeing conditions while maximizing field of view within the detector and space limitations. The optical configuration prioritizes compactness, manufacturability, and alignment robustness, ensuring compatibility with the telescope interface and long-term operational stability.

    \subsection{Telescope}\label{subsec:Telescopio}
    
    The host facility is the 2.15\,m Jorge Sahade telescope, located in the province of San Juan, Argentina. The telescope operates in a Cassegrain configuration with an effective focal ratio of $f/8.5$. Its main optical parameters are summarized in Table\,\ref{tab:OP_JS_telescope}.
    
    \begin{table}[ht]
    \vspace{-2mm}
    \caption{Main optical parameters of the Jorge Sahade telescope.}
    \label{tab:OP_JS_telescope}
    \centering
    \small
    \begin{tabular}{c c c c}
    \toprule
    Radius of curvature & Separation & Semi-diameter & Conic\\
    (mm) & (mm) & (mm) & constant\\
    \midrule
    $-1.118\times10^{4}$ & $4052.6$ & $1076.0$ & $-1.07$\\
    $-4300$              & $951.8$  & $317.5$  & $-4.32$\\
    \bottomrule
    \end{tabular}
    \end{table}
    
    The effective focal length of the JS telescope is $18\,213.6$\,mm, which for a detector format of $22.5\,\times\,22.5$\,mm corresponds to an angular field of $4.2\,\arcmin\,\times\,4.2\,\arcmin$ at the Cassegrain focus. The instrument is directly coupled to this focus, preserving the telescope optical geometry while operating under the mechanical and operational constraints imposed by the telescope structure, whose quantitative formulation and implications for the optical design are discussed in Section\,\ref{subsec:SpecIntm}.
    
    \begin{figure}[ht]
    \centering
    \includegraphics[width=0.9\columnwidth]{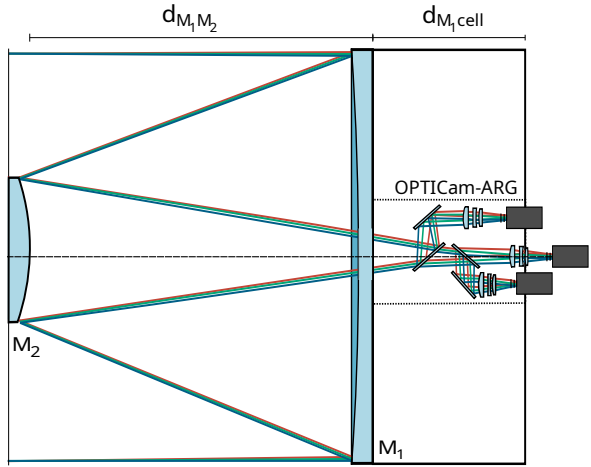}
    \caption{\justifying General layout of the Jorge Sahade (JS) telescope in Cassegrain configuration ($f/8.5$) coupled to the \textit{OPTICAM-ARG} instrument. The main optical elements of both the telescope and the instrument are shown, together with the geometric constraints imposed by the telescope cell. The available length within the cell, $d_{\mathrm{M_1cell}}$, limits the axial extension of the instrument, while the admissible mechanical volume is indicated by a dashed line.}
    \label{fig:CompleteSystem}
    \end{figure}
    
    Figure\,\ref{fig:CompleteSystem} shows the general layout of the JS telescope coupled to \textit{OPTICAM-ARG}. In this nomenclature, M1 and M2 denote the primary and secondary mirrors of the telescope, respectively, while mirrors introduced by the instrument are labeled sequentially. The distance between the primary mirror vertex and the image plane is $951.8$\,mm (Table\,\ref{tab:OP_JS_telescope}), whereas the axial length available within the telescope cell is $d_{\mathrm{M}_1\mathrm{cell}}\,=\,703.15$\,mm (Figure\,\ref{fig:CompleteSystem}). Consequently, only $248.65$\,mm remain available beyond the cell boundary for instrument integration, which motivates full accommodation of the instrument within the telescope cell. The admissible mechanical volume therefore constrains both the axial extension and the maximum diameter of the optical train.

\begin{figure*}[t]
    \centering
    \includegraphics[width=0.65\textwidth]{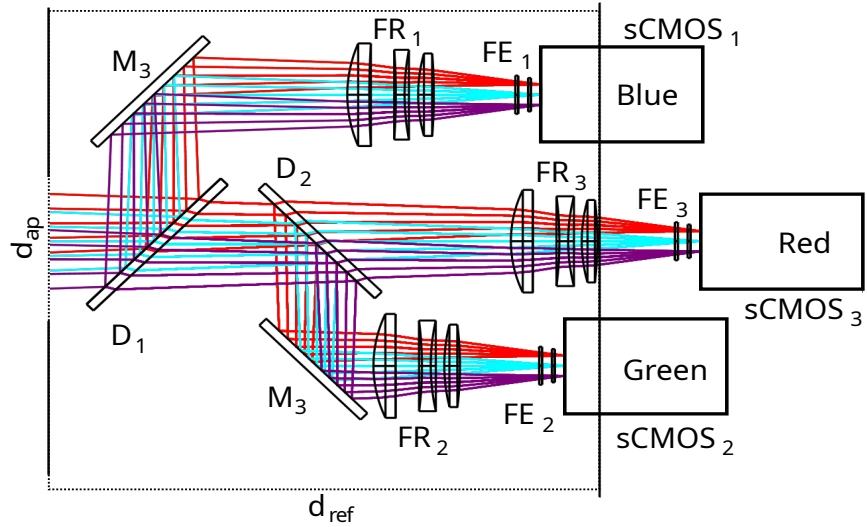}
    \caption{\justifying Enlarged view of the instrument within the mechanical volume available inside the JS telescope cell. The axial reference from the primary mirror vertex, the distance to the mounting flange $d_{\mathrm{ref}}$, and the mechanical aperture diameter $d_{\mathrm{ap}}$ are indicated. The available integration volume is shown by a dotted line, while the camera volumes are represented by solid rectangles. Two fold mirrors, denoted as $\mathrm{M}_3$, are included to accommodate the beam geometry within the mechanical constraints. The main optical elements of each channel are also identified: dichroics ($\mathrm{D}_i$), focal reducers ($\mathrm{FR}_i$), filter assemblies ($\mathrm{FE}_i$), and sCMOS detectors ($\mathrm{sCMOS}_i$). The three spectral channels are labeled as Blue, Green, and Red directly on the detector elements.}
    \label{fig:Tri-channel}
    \end{figure*}

    An enlarged view of the instrument within these constraints is presented in Figure\,\ref{fig:Tri-channel}. An axial reference located $140\,\mathrm{mm}$ from the primary mirror vertex is introduced, together with a reference distance to the mounting flange of $\mathrm{d}_{\mathrm{ref}}\,=\,563.15\,\mathrm{mm}$ and a mechanical aperture diameter of $\mathrm{d}_{\mathrm{ap}}\,=\,508\,\mathrm{mm}$. The admissible integration volume is indicated by a dotted line. The volumes occupied by the cameras are also shown as solid rectangles of $165.8\,\times\,103.0\,\mathrm{mm}$, providing a geometric reference for evaluating mechanical compatibility within the available integration volume. Under these geometric constraints, the optical design is developed from the converging beam delivered by the telescope at the Cassegrain focus.
    
    \subsection{Dichroics}\label{subsec:Dichroics}
    
    Spectral separation is achieved by two dichroics operating in the converging beam, defining three channels that cover the blue, green, and red regions of the optical spectrum. The adopted channel separation and associated filter allocation are summarized in Table\,\ref{tab:Channel_separation}.
    
    \begin{table}[ht]
    \vspace{-2mm}
    \caption{Spectral separation of the channels and associated filters.}
    \label{tab:Channel_separation}
    \centering
    \begin{tabular*}{0.9\columnwidth}{@{}>{\centering\arraybackslash}p{0.2\columnwidth}
                           >{\centering\arraybackslash}p{0.4\columnwidth}
                           >{\centering\arraybackslash}p{0.3\columnwidth}@{}}
    \toprule
    Channel & Spectral range & Filters\\
    \midrule
    Blue  & $0.35$--$0.55\,\mu$m & \textit{u'}, \textit{g'}\\
    Green & $0.55$--$0.67\,\mu$m & \textit{r'}\\
    Red   & $0.67$--$1.00\,\mu$m & \textit{i'}, \textit{z'}\\
    \bottomrule
    \end{tabular*}
    \end{table}
    
    The incident beam first reaches dichroic $\mathrm{D}_1$, which reflects the blue channel over $0.35$--$0.55\,\mu$m and transmits the remaining spectral range. The transmitted beam then reaches $\mathrm{D}_2$, which reflects the green channel ($0.55$--$0.67\,\mu$m) and transmits the red channel ($0.67$--$1.00\,\mu$m), as illustrated in Figure\,\ref{fig:Tri-channel}. In this context, the dichroic elements are defined in terms of their spectral separation bands, while the detailed coating design is not included at this stage, and the spectral ranges are therefore treated as idealized boundaries for the optical design.
    
    The reflected blue channel beam propagates through its focal reduction system ($\mathrm{FR}_1$; Section\,\ref{subsec:FocalReducer}), the interchangeable filter assembly ($\mathrm{FE}_1$; Section\,\ref{subsec:Filter}), and finally reaches its detector ($\mathrm{sCMOS}_1$; Section\,\ref{subsec:sCMOS}). This arm incorporates a flat mirror ($\mathrm{M}_3$) to adapt the beam path to the mechanical constraints imposed by integration within the telescope cell.

    The green channel follows an analogous sequence through its focal reducer ($\mathrm{FR}_2$), filter assembly ($\mathrm{FE}_2$), and detector ($\mathrm{sCMOS}_2$), also incorporating a flat folding mirror to accommodate the available mechanical volume. In contrast, the red channel propagates directly toward its focal reducer ($\mathrm{FR}_3$), filter assembly ($\mathrm{FE}_3$), and detector ($\mathrm{sCMOS}_3$), without requiring additional beam folding.
    
    Because the dichroics operate in a converging beam, their insertion breaks the rotational symmetry of the optical system and introduces off-axis aberrations, primarily coma and astigmatism. To mitigate these effects, a wedge is introduced on the second surface of each dichroic \citep{Howard:85}, whose optimization is described in Section\,\ref{subsec:NumericDesign}. Once spectral separation is established, each channel propagates independently along its optical arm, where the design concentrates on achieving the required plate scale and controlling the residual aberrations introduced by focal reduction through the dedicated focal reduction systems described in Section\,\ref{subsec:FocalReducer}.

    \subsection{Focal reducer}\label{subsec:FocalReducer}
    
    The focal reduction system $\mathrm{FR}_i$ in each channel consists of three lenses ($L_1$, $L_2$, and $L_3$), reducing the telescope's effective focal length to approximately $9\,127.2$\,mm. The configuration is determined through optimization of the lens radii of curvature, together with the separation between the third lens and the filter plane, $d_{FR_iFE_i}$.
    
    Although the initial design was based on the $\mathrm{SFPL51}$ and $\mathrm{F2HT}$ glasses used in \textit{OPTICAM}, the final design adopts $\mathrm{K\text{-}PFK85}$\footnote{Sumita Optical Glass, product catalog available at \url{https://www.sumita-opt.co.jp/en/products/optical.html}.} for the first and third lenses and $\mathrm{ADF355}$\footnote{HOYA Corporation, optical materials catalog available at \url{https://www.hoya.com/en/}.} for the central lens. This choice is driven by improved chromatic balance and enhanced concentration of polychromatic energy. The use of a three-lens focal reducer per channel is adopted as a design condition derived from the instrumental specifications. This choice balances the minimization of optical surfaces with the need to maximize the corrected field on the detector while satisfying plate scale requirements, site seeing conditions, and geometric constraints. In addition, an effective lens diameter limit of $\sim100$\,mm is imposed to ensure mechanical feasibility, availability of commercial optical materials, and practical manufacturability and assembly. The quantitative justification for these criteria is presented in Section\,\ref{subsec:SpecIntm}.

    As a result, the angular field projected onto the sCMOS detector increases from the $4.2\,\arcmin\,\times\,4.2\,\arcmin$ field at the telescope focus (Section\,\ref{subsec:Telescopio}) to $8.4\,\arcmin\,\times\,8.4\,\arcmin$, while maintaining a uniform plate scale of $22.6\,\arcsec\,\mathrm{mm}^{-1}$ across the three channels. Once the focal reduction is established, the beam in each channel propagates toward the filter stage.
    
    \subsection{Filters and windows}\label{subsec:Filter}
    
    The SDSS\footnote{\url{https://www.sdss.org/}} (\textit{Sloan Digital Sky Survey}) photometric system is adopted as the primary filter set, as it constitutes a standard reference in observational astronomy and its spectral properties are well characterized. Originally described by \citet{1996AJ....111.1748F}, this system comprises five bands (\textit{u'}, \textit{g'}, \textit{r'}, \textit{i'}, and \textit{z'}) that provide nearly continuous coverage from the atmospheric cutoff in the near ultraviolet ($\sim0.3\,\mu$m) to the sensitivity limit of CCD and sCMOS detectors in the near infrared ($\sim1.0\,\mu$m).
    
    A key advantage of the SDSS system is the minimal overlap among its transmission bands. This feature is particularly suitable for multichannel instruments such as \textit{ULTRACAM }\citep{2007MNRAS.378..825D} and \textit{OPTICAM}, as it reduces distortions of the effective system response near dichroic cutoffs. In contrast, classical systems such as Johnson–Morgan–Cousins (\textit{U, B, V, R$_\mathrm{C}$, I$_\mathrm{C}$}) exhibit substantial band overlap, complicating their joint implementation with dichroics and altering the effective spectral response of each channel.
    
\begin{figure}[ht]
    \centering
    \includegraphics[width=1.0\columnwidth]{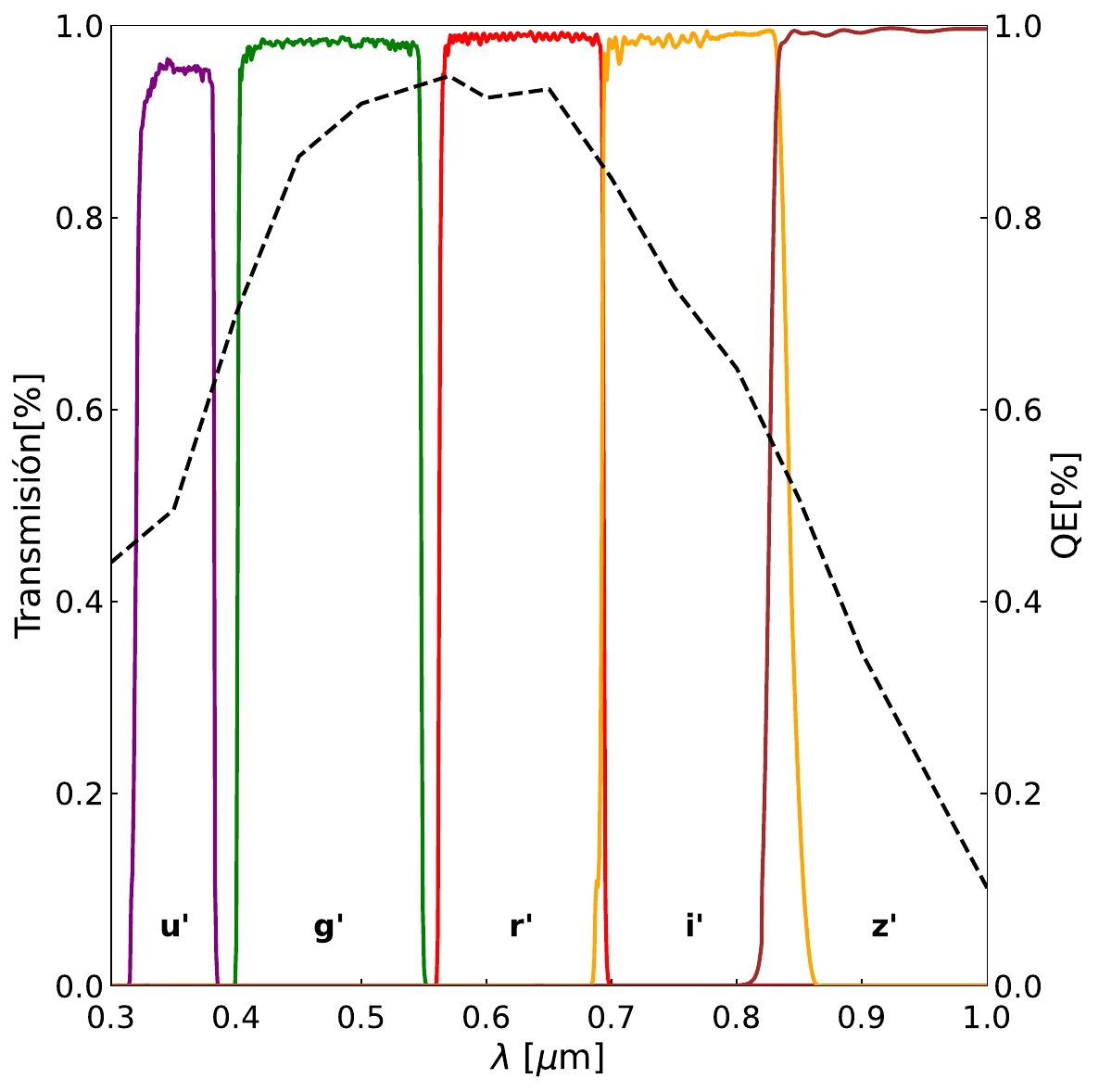}
    \caption{\justifying Transmission profiles of the Astrodon Gen2 SDSS filter set ($u' g' r' i' z'$; colored curves) and the quantum efficiency (QE) curve of the Andor Marana\,4.2B--11 sCMOS detector (black dashed line).}
    \label{fig:Filters}
    \end{figure}

    In the \textit{OPTICAM-ARG} configuration, each optical arm allows the use of interchangeable SDSS filters. The transmission curves of the Astrodon Gen2 SDSS filter set were adopted as reference profiles for the optical design. Such curves are shown in Figure\,\ref{fig:Filters}. The same figure includes the quantum efficiency (QE) curve of the Andor Marana\,4.2B--11 sCMOS detector (see Section\,\ref{subsec:sCMOS}), enabling assessment of the combined spectral response of the optical system and detector.
    
    The filter allocation per channel, summarized in Table\,\ref{tab:Channel_separation}, is as follows: the blue channel employs \textit{u'} and \textit{g'}, the green channel uses \textit{r'}, and the red channel incorporates \textit{i'} and \textit{z'}. This distribution provides balanced spectral coverage while maintaining compatibility with the dichroic cutoffs and detector response. The final optical elements before detection are the entrance detector windows, which provide optical coupling to the sCMOS sensor while preserving the vacuum seal of the camera head. The windows are plane parallel elements made of fused silica, selected for high transmission across the operational range and low birefringence.
    
    \subsection{Detectors}\label{subsec:sCMOS}
    
    \textit{OPTICAM-ARG} is designed around the Andor Marana\,4.2B--11 sCMOS detector in each channel. The Marana\,4.2B--11 detector is based on back-illuminated sCMOS technology and provides quantum efficiencies of up to $95\%$ in the visible range (Figure\,\ref{fig:Filters}), while supporting high-cadence applications. The sensor is a $2048\,\times\,2048$ array of $11\,\mu$m square pixels, corresponding to an effective diagonal of approximately $32$\,mm. These dimensions define the sensitive area and constitute a boundary condition for the field of view and plate scale adopted in the optical design. According to the manufacturer specifications, full-frame readout can be achieved in less than $50$\,ms \citep{andor_catalogo}, enabling exposure times from microseconds to several seconds. Marana\,4.2B--11 incorporates a vacuum-cooled system that enables operation down to $-45^\circ$C, thereby reducing thermal noise.
    
    The optical configuration described in this section, together with the telescope's geometric constraints and the adopted detector model, defines the boundary conditions for the design process. The following section presents the methodology used to construct, optimize, and refine the optical design.
        
\section{Design methodology}\label{sec:Metodologia}

    The optical design is developed within a structured framework that combines geometrical optics principles with numerical optimization techniques (Sections\,\ref{subsec:SPD} and\,\ref{subsec:NumericDesign}), aiming to construct physically consistent and reproducible solutions from the earliest stages. This methodology builds upon the classical foundations established by Conrady and Kingslake \citep{conrady1957applied, conrady1960applied, kingslake1978lens} and adopts a staged workflow similar to that described by \citet{bentley2012field}, in which model complexity increases progressively under controlled conditions.
            
\begin{figure}[ht]
        \centering
        \includegraphics[width=0.8\columnwidth]{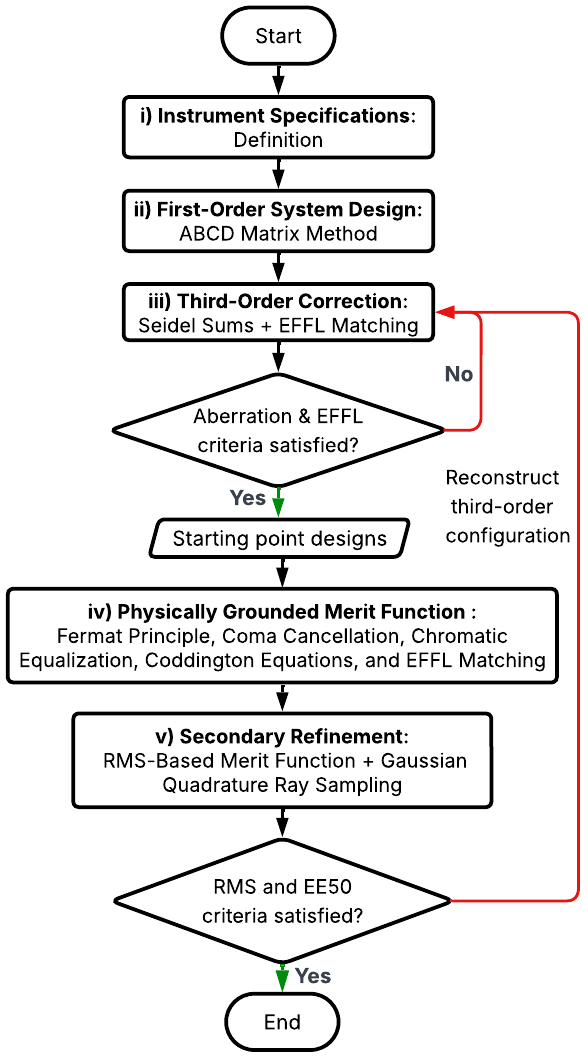}
        \caption{\justifying Flow diagram of the adopted optical design methodology. The process advances through successive design stages, starting from the definition of instrumental specifications, followed by first-order design using ABCD matrices, third-order correction based on Seidel sums and effective focal length adjustment, numerical optimization through a physically grounded merit function, and a final geometric refinement based on RMS spot radius minimization using Gaussian quadrature sampling.}
        \label{fig:OverallFC}
    \end{figure}

    Figure\,\ref{fig:OverallFC} summarizes the design sequence, from the definition of instrumental specifications to the final refinement through exact ray tracing, and provides the conceptual guide for the subsections that follow.
    
    \subsection{Starting point designs}\label{subsec:SPD}

    The starting point designs of \textit{OPTICAM-ARG} are constructed through a progressive formulation of the optical system that integrates instrumental specifications (Section\,\ref{subsec:SpecIntm}) with geometrical optics modeling, including paraxial analysis (Section\,\ref{subsec:FirstOrder}) and third-order corrections (Section\,\ref{subsec:ThirdOrder}). This staged construction yields a physically coherent initial configuration that serves as the basis for subsequent numerical refinement. The following subsubsections detail this process, beginning with the quantitative definition of the instrumental constraints and proceeding to first and third-order modeling.

        \subsubsection{Instrument specifications}\label{subsec:SpecIntm}
        
        The instrumental specifications of the system are formulated quantitatively in order to translate the scientific requirements and the constraints imposed by the telescope into optical parameters directly usable in the design process. These specifications guide the definition of the overall system geometry, the selection of optical elements, and the choice of the spatial scales relevant for image formation. As discussed in Section\,\ref{subsec:Telescopio}, the instrument is designed to operate at the Cassegrain focus of the JS telescope, with an effective focal length of $18\,213.6$\,mm and a focal ratio of $f/8.5$. In that section, it was established that the position of the image plane with respect to the primary mirror cell imposes an axial constraint, given that the distance between the primary mirror vertex and the image plane is $951.8$\,mm, while the axial length occupied by the cell is $d_{\mathrm{M}_1\mathrm{cell}}\,=\,703.15$\,mm. As a consequence, the image plane is located approximately $248.65$\,mm beyond the boundary of the telescope cell.
                            
        This separation is insufficient to externally accommodate the instrument elements required for multichannel operation, including beam splitters, the focal reduction system, filters, and detectors. Consequently, instrument integration must be carried out predominantly within the telescope cell volume. Under this condition, the axial and transverse arrangement of the elements is constrained, since additional components would require moving closer to the primary mirror vertex, where the beam is more convergent, and its diameter increases significantly.
                            
        When operating within the cell, excessive proximity to the primary mirror vertex must be avoided, as this would lead to lens sizes incompatible with mechanical feasibility, the availability of commercial optical glasses, and manufacturing and assembly processes. In this context, an additional constraint is imposed such that the effective lens diameter does not exceed $\sim100$\,mm. Complementarily, the maximum diameter allowed for mechanical integration of the instrument is limited by the available aperture in the telescope cell, defined by $\mathrm{d}_{\mathrm{ap}}\,=\,508$\,mm.
                            
        A detailed representation of these geometric and mechanical constraints is shown in Figure\,\ref{fig:Tri-channel}. In this figure, a reference is introduced at $140$\,mm from the primary mirror vertex, together with a reference distance to the mounting flange of $\mathrm{d}_{\mathrm{ref}}\,=\,563.15$\,mm. Under this geometric convention, the first instrument element, corresponding to the first surface of the first dichroic, is located at a distance of $247.43$\,mm measured from the primary mirror vertex. The camera volumes are also included, allowing verification of the mechanical compatibility of the adopted optical layout, particularly the beam folding achieved through $45^\circ$ flat mirrors in the blue and green channels.

        \begin{table*}[hb]
            \centering
            \caption{Adopted instrumental specifications and design constraints.}
            \label{tab:InstrumentSpecs}
            \begin{tabular}{l c p{5.5cm}}
                \toprule
                \textbf{Parameter} & \textbf{Value} & \textbf{Comment} \\
                \midrule
                Telescope & JS 2.15 m ($f/8.5$) & Cassegrain configuration \\
                Telescope focal length & $18\,213.6$ mm & Nominal configuration \\
                Target focal length & $9\,127.2$ mm & Per-channel focal reduction \\
                Plate scale & $22.6\,\arcsec\,\mathrm{mm}^{-1}$ & Uniform across all channels \\
                Field of view per channel & $8.4\arcmin \times 8.4\arcmin$ & Photometry \\
                Pixel size & $11\,\mu$m & Marana 4.2B--11 sCMOS \\
                Angular sampling & $\sim 0.25\,\arcsec\,\mathrm{pix}^{-1}$
                & Oversampled relative to site seeing \\
                Typical site seeing & $\sim1.5\,\arcsec$ & DIMM measurements \\
                Available axial volume & $248.65$ mm & From the image plane \\
                Maximum mechanical aperture & $508$ mm & Telescope cell \\
                Maximum lens diameter & $\sim100$ mm & Mechanical feasibility \\
                Minimum number of lenses & 3 per channel & Field and chromatic correction \\
                Spectral range & $0.35$--$1.00\,\mu$m & Simultaneous operation \\
                \bottomrule
            \end{tabular}
        \end{table*}
            
         The joint definition of the field of view and the plate scale constitutes a central aspect of the instrumental specifications, as both parameters directly determine the feasibility of performing differential photometry in high-cadence observations. In particular, the field of view must be sufficiently large to ensure a high probability of including suitable comparison stars within the same frame. Following the approach adopted in ULTRACAM, a search radius of approximately $5\,\arcmin$ provides an $\sim80\%$ probability of finding a comparison star of magnitude $R\,=\,12$ at intermediate Galactic latitudes \citep{simons1995technical}. Consequently, adopting a field of view larger than $5\,\arcmin$ effectively guarantees the availability of reference stars.

        The plate scale is defined following the sampling criterion implemented in \textit{OPTICam}, where adequate image sampling under typical seeing conditions was prioritized while maximizing detector usage. For \textit{OPTICam-ARG}, the adopted configuration yields a plate scale of $22.6\,\arcsec\,\mathrm{mm}^{-1}$, which corresponds to focal reduction of the JS telescope to an effective focal length of approximately $9\,127$\,mm per channel. For a $22.5$\,mm image-plane format, this configuration provides an angular field of approximately $8.4\,\arcmin \times 8.4\,\arcmin$.

        The seeing conditions at the JS Telescope site have been characterized through DIMM measurements, reporting values between $1.31\,\arcsec$ and $2.03\,\arcsec$, with a representative value of $1.54\,\arcsec$ \citep{mammanamediciones}. Under these conditions, the Nyquist sampling criterion requires an angular pixel scale of approximately $\lesssim 0.75\,\arcsec\,\mathrm{pix}^{-1}$. With a detector pixel size of $11\,\mu$m, the adopted plate scale corresponds to an angular sampling of $\sim0.25\,\arcsec\,\mathrm{pix}^{-1}$, which is well below this limit and therefore represents an oversampled regime with respect to the site seeing. Consequently, the system response, when expressed in terms of an equivalent full width at half maximum (FWHM) derived from EE50, remains well sampled relative to the atmospheric seeing.

        As part of the definition process for the focal reduction system, an alternative configuration composed of two lenses was preliminarily evaluated to minimize the number of optical surfaces. However, this solution could not provide adequate correction of the required field and plate scale, resulting in spot diagrams with radii of order $324.26\,\mu$m, incompatible with the detector pixel size and the site seeing conditions. Consequently, the use of three lenses per channel was adopted as a minimum design condition, establishing a compromise between reducing optical surfaces and maximizing the corrected field of view on the detector.
        
        Table\,\ref{tab:InstrumentSpecs} consolidates the instrumental specifications and geometric constraints that define the boundary conditions of the optical design. These parameters establish the operational and structural framework within which the subsequent modeling and optimization stages are carried out.
        
        The selection of optical materials constitutes a key instrument specification, given its direct impact on the correction of chromatic aberrations over the considered spectral interval. In particular, combinations of low and high dispersion materials are prioritized. As an initial reference, the $\mathrm{S\text{-}FPL51}$ and $\mathrm{F2HT}$ glasses previously employed in \textit{OPTICAM} are adopted. Based on this starting point, a local material space in the $(n, V_d)$ plane is explored, imposing extensions of $\Delta n\,=\,0.025$, $\Delta V_d\,=\,5.0$, and a minimum spectral transmission threshold of $PT_{\mathrm{threshold}}\,=\,0.8$. As a result, the $\mathrm{K\text{-}PFK85}$ and $\mathrm{ADF355}$ glasses are selected for the final design of the focal reduction system. These materials are fixed at this stage and are subsequently used in the construction of the initial optical model, whose development continues in the following subsection under the paraxial approximation.          
                    
        \subsubsection{First-order design}\label{subsec:FirstOrder}
     
        The first-order design is formulated using the paraxial approximation and the ABCD matrix formalism \citep{10.1119/1.1970159}. Within this framework, the optical system is described in terms of the optical powers of the lenses and their separations under the thin lens approximation. The element separations are kept fixed, and the solution space is restricted to determining the powers of the three lenses, thereby establishing an initial description of the global optical behavior.
                
        As an initial condition for assigning the signs of the optical powers, the first solution of the Cooke triplet is adopted \citep[Fig.\,192]{conrady1960applied}, characterized by positive power outer lenses and a negative power central lens. This configuration leads to powers of comparable magnitude and provides a balanced distribution suitable for subsequent refinement stages. This solution is obtained by imposing three independent constraints: (i) agreement between the target effective focal length, $\mathrm{EFFL}_{\mathrm{t}}$, and the effective focal length of the paraxial system, $\mathrm{EFFL}_{\mathrm{T}}$; (ii) the paraxial focusing condition, expressed through the $d$ term of the global ABCD matrix; and (iii) a regularization term that penalizes extreme power values. These quantities are defined as
                
        \begin{equation}
        \label{eq:Meritfunparax}
        \begin{split}
        d\phi\,&=\,\sum_{i=1}^{3}\frac{\phi_{\mathrm{L}i}}{n_{\mathrm{L}i}}, \\
        d\mathrm{EFFL}\,&=\,\left| \mathrm{EFFL}_{\mathrm{t}} - \mathrm{EFFL}_{\mathrm{T}} \right|, \\
        d\,&=\,\mathrm{MS}_{\mathrm{OPTICAM\text{-}ARG}}[1,1],
        \end{split}
        \end{equation}
                                
        where $\mathrm{MS}_{\mathrm{OPTICAM\text{-}ARG}}$ denotes the paraxial transfer matrix of the complete system.
                
        Once the thin lens paraxial solution is obtained, the model is extended to a thick lens representation for subsequent stages of the design.

        \subsubsection{Third-order design}\label{subsec:ThirdOrder}
    
        The third-order formulation describes the optical system through physically realizable geometric parameters associated with thick lenses. The degrees of freedom of the design are defined primarily by the radii of curvature of each surface and by an axial compensation distance, while the optical materials are kept fixed in accordance with the instrumental specifications previously established (Section\,\ref{subsec:SpecIntm}).

        The lens thicknesses are fixed following an empirical manufacturability criterion, based on the principle described by \citet{kingslake1978lens} and refined by \citet{bentley2012field}, which states that an approximate diameter to thickness ratio between $10{:}1$ and $5{:}1$ facilitates fabrication.

        The initial radii of curvature are derived from the lensmaker equation under the constraint $\mathrm{R}_{2i}\,=\,\beta\,\mathrm{R}_{1i}$, which preserves the prescribed optical power while defining a specific geometric family of solutions. This structural assumption constrains the admissible region of the design space and directly influences the numerical conditioning and convergence landscape of the subsequent optimization stages. Within this context, the optimization is formulated as the minimization of third-order aberrations through the Seidel sums, explicitly expressed as functions of the radii of curvature. The five classical monochromatic coefficients are considered: spherical aberration ($S_{I}$), coma ($S_{II}$), astigmatism ($S_{III}$), field curvature ($S_{IV}$), and distortion ($S_{V}$), following the formalism of \citet{kidger2001fundamental} and \citet{welford2017aberrations}. In addition, longitudinal chromatic aberration ($C_{L}$) is included, evaluated at three representative wavelengths for each spectral channel.
                    
        The evaluation of the Seidel sums is performed using the optical simulation library \texttt{KrakenOS} \citep{herrera2022krakenos} \footnote{\url{https://github.com/Garchupiter/Kraken-Optical-Simulator}}, through the \texttt{SeidelTool}, which allows the computation of both individual surface contributions and the total monochromatic and chromatic sums of the system. Under this formulation, the optimization problem is defined by seven design variables: the six radii of curvature of the focal reduction system and an axial compensation distance $d_{FR_iFE_i}$, introduced to adjust the effective focal length. 
        
        The merit function is constructed from the five monochromatic Seidel terms, the longitudinal chromatic term, and a penalty term associated with the effective focal length, and is formulated as
        
        \begin{equation}
        \min_{\mathbf{x}}\;\frac{1}{2}\sum_{i = 1}^{7} \rho\left(f_i(\mathbf{x})^2\right),\,
        \mathbf{lb}\leq\mathbf{x}\leq\mathbf{ub},
        \label{eq:LSeq}
        \end{equation}
        where $\mathbf{x}\,\in\,\mathbb{R}^7$ is the vector of design variables, with $\mathbf{lb}$ and $\mathbf{ub}$ defining the corresponding lower and upper bounds, $f_i$ the residual components, and $\rho$ a robust loss function. Explicitly, the residual vector is expressed as
        
        \begin{equation}
        \mathbf{f}(\mathbf{x})^{\mathsf{T}}\,=\,\bigl(
        S_{I},
        S_{II},
        S_{III},
        S_{IV},
        S_{V}^{*},
        C_{L},
        \Delta \mathrm{EFFL}
        \bigr),
        \label{eq:merit_vector}
        \end{equation}
                    
        where $\Delta \mathrm{EFFL}$ denotes the difference between the required effective focal length and that obtained from the optical model.
                                    
        The optimization is solved using a nonlinear least-squares scheme implemented in \texttt{SciPy} \citep{2020SciPy-NMeth}, starting from the initial values obtained in the paraxial model. Bounds are imposed on each variable based on geometric considerations and mechanical feasibility.
                                    
        Convergence of the Seidel based merit function does not automatically imply acceptance of the solution as a definitive initial configuration. Instead, the third-order stage functions as a conditional validation step within the overall design sequence. Only when both the aberration balance and the effective focal length criteria are simultaneously satisfied is the resulting configuration adopted for the numerical optimization stages. Otherwise, the process returns to the third-order stage, where a new geometric configuration is reconstructed under the imposed $\beta$ constraint, while preserving the paraxial power distribution previously established.

    \subsection{Numeric Design}\label{subsec:NumericDesign}

    The numerical design is organized into two complementary stages implemented in the \texttt{KrakenOS} environment: (a) a physically grounded optimization, in which a physically grounded merit function (PGMF) is applied for numerical refinement starting from the validated third-order configuration; and (b) a geometric refinement aimed at improving image quality through minimization of the RMS spot radius while preserving the structural coherence of the validated configuration. Both stages rely on exact ray tracing for performance evaluation.

    In the first stage (Section\,\ref{subsec:PGMF}), the merit function terms enable direct quantification of the dominant aberrations and promote numerically stable behavior consistent with the physical interpretation established in the preceding stages. The second stage (Section\,\ref{subsec:RMS}) performs a geometric refinement by minimizing the RMS spot radius using Gaussian quadrature pupil sampling. This refinement operates on a configuration that has been physically stabilized, enabling improvement of image quality without perturbing the global optical structure or reintroducing ill-conditioned regions into the convergence landscape.
                            
    As part of this second stage, and based on the design optimized in \texttt{KrakenOS}, the commercial software ZEMAX\textsuperscript{\textregistered} OpticStudio \footnote{\url{https://www.ansys.com/products/optics/ansys-zemax-opticstudio}} is used as an independent verification and fine-tuning tool. In particular, it is employed to determine the system's best focus position and to optimize the wedge angle of the second surface of the dichroics.
    
    All acceptance criteria governing the design framework, including effective focal length matching, aberration balance, and RMS and EE50 performance, are evaluated within the \texttt{KrakenOS} environment. ZEMAX\textsuperscript{\textregistered} is used for independent verification and final geometric adjustment, and it does not modify the conditional logic that triggers reconstruction of the initial configuration.
    
        \subsubsection{Physically grounded optimization}\label{subsec:PGMF}
            
        The first stage corresponds to the application of the PGMF for the numerical refinement of the optical design. The numerical evaluation of the terms composing the PGMF is carried out directly using \texttt{KrakenOS}, which provides the ray trajectories, total optical path lengths, and intersection positions required to quantify each of the considered aberrations.
                                    
        The correction of spherical and chromatic aberrations is grounded in Fermat's principle, through the evaluation of variations in the total optical path length for different ray types and wavelengths. Spherical aberration is controlled by imposing that the difference $\delta l$ between the optical paths corresponding to the marginal ray and the chief ray tends toward zero. In turn, chromatic correction is quantified from the variation $\delta r$ of the total optical path, evaluated at three representative wavelengths per spectral channel \citep{malacara2003handbook}.
                                                
        Coma aberration is reduced through direct minimization of transverse coma, quantified from the difference in intersection height at the image plane between two marginal rays and the chief ray \citep{conrady1957applied}. Astigmatism is evaluated from the separation between the sagittal and tangential foci, obtained through differential ray tracing near the optical axis, in accordance with the formulation of the Coddington equations \citep{kingslake1978lens}. As an additional constraint, the difference between the required effective focal length and that obtained from the optical model, $\Delta \mathrm{EFFL}$, is incorporated. Since coma and astigmatism are off-axis aberrations, these terms are evaluated at two field points located in extreme regions of the field of view.
                                                
        The PGMF is constructed as a residual vector that includes contributions associated with spherical aberration, chromatic aberration, coma, astigmatism, and effective focal length deviation. The optimization is solved using a nonlinear least squares scheme implemented in \texttt{SciPy}, while keeping the lens thicknesses and non critical separations fixed and restricting the design variables to physically consistent intervals. This stage provides a numerically stable and physically coherent design, which serves as the starting point for the geometric refinement stage described below.
                  
        \subsubsection{Subsequent RMS-based refinement}\label{subsec:RMS}
                
        The refinement stage is devoted to improving the system image quality through minimization of the RMS spot radius. This procedure reduces residual aberrations not fully captured by the physically grounded analysis, while preserving the structural coherence of the validated configuration. The RMS metric is computed using a Gaussian quadrature pupil sampling scheme, enabling efficient and accurate evaluation of geometric performance through weighted sampling of the circular aperture.
                                                        
        Starting from the design-variable vector obtained in the previous stage, the lens radii of curvature are optimized using \texttt{KrakenOS} in conjunction with \texttt{SciPy}, adopting the RMS spot radius as the objective metric. This metric is evaluated over six field points defined by symmetry and at four wavelengths per spectral channel. The optimized parameters are subsequently transferred to ZEMAX\textsuperscript{\textregistered} OpticStudio, where the system best-focus position is determined.
                                                        
        As discussed in Section\,\ref{subsec:Dichroics}, the wedge angles of the second surface of the dichroics are then optimized in ZEMAX\textsuperscript{\textregistered} to compensate for off-axis aberrations introduced by the converging beam. This adjustment is performed by minimizing the RMS spot radius over a set of twelve field points representative of the effective rectangular aperture of each dichroic. A final fine adjustment of the lens curvatures is then carried out.
                                                        
        As the final component of the numerical design, this refinement stage also functions as a global validation layer within the overall workflow. If the RMS and the radius enclosing $50\%$ of the total energy (EE50) criteria are not simultaneously satisfied after geometric optimization, the methodology prescribes reconstruction of a new initial configuration at the third-order level, thereby reinitializing the iterative process. In this way, convergence is not forced within an ill-conditioned local region of the design space, but instead achieved through controlled updates of the geometric model and reconditioning of the convergence landscape. This reconstruction logic is implemented within the open-source computational framework, ensuring that convergence control remains independent of proprietary optimization environments.

\section{Results}\label{sec:Resultados}

This section presents the results of the final optical design of \textit{OPTICAM-ARG}, after incorporation of dichroic wedging and completion of the final adjustment of the focal reducers. The nominal optical performance and the sensitivity of the system to manufacturing tolerances are analyzed. The complete optical prescription, including the three focal reducers and dichroic elements, is summarized in Table\,\ref{tab:optical_prescription}.

Nominal performance is assessed using image quality metrics based on exact ray tracing, including the RMS spot radius, the geometric (GEO) spot radius, and EE50. The analysis is carried out at three representative field points: the on-axis field $(0.00,0.00)$ and two off-axis positions, $(0.00,0.07)$ and $(0.07,0.07)$, consistently adopted for all spectral channels. The corresponding spot diagrams and encircled energy curves are shown in Figures\,\ref{fig:SPT_all} and\,\ref{fig:EE_All}, and the quantitative metrics are summarized in Table\,\ref{tab:NominalMetrics} (Section\,\ref{subsec:Nominal}). This field selection captures both the near axis image quality and the off-axis behavior across the field of view.

The optimized dichroic wedging configuration, adopted to mitigate off-axis aberrations, is described and quantified in Section\,\ref{subsec:Dicacu}. Manufacturing sensitivity is subsequently evaluated through a tolerance analysis (Section\,\ref{subsec:TolAnalysis}) that considers perturbations in both lenses and dichroics. The adopted tolerance ranges are listed in Table\,\ref{tab:TolerancesGlobal}, and the resulting degradation relative to the nominal design is expressed as an equivalent FWHM including atmospheric seeing. The final FWHM budget is summarized in Table\,\ref{tab:performance_canales_total}.

\begin{table*}[!ht]
\centering
\caption{\justifying Optical prescription of the \textit{OPTICAM-ARG} instrument. The wedge angles represent the effective tilt values provided by ZEMAX\textsuperscript{\textregistered} under the tilted-surface representation of the dichroic elements.} 

\label{tab:optical_prescription}
\begin{tabular}{l c c c}
\toprule
\textbf{Parameter} & \textbf{Channel 1 (Blue)} & \textbf{Channel 2 (Green)} & \textbf{Channel 3 (Red)} \\
\midrule
\multicolumn{4}{c}{\textit{Focal Reducer}} \\
\midrule
Lens 1 radius $R_{1}$ (mm)          & 146.68 & 137.65 & 134.86 \\
Lens 1 radius $R_{2}$ (mm)          & 973.07 & 1.68E+4 & -1.05E+4 \\
Lens 1 thickness $t_{1}$ (mm)       & 22.50  &  22.50  & 22.50 \\
Lens 1 glass                        &$\mathrm{K\text{-}PFK85}$&$\mathrm{K\text{-}PFK85}$& $\mathrm{K\text{-}PFK85}$\\
Lens 1 Semi-Diameter (mm)                & 56.00 & 56.00 & 56.00 \\
Lens separation $d_{12}$ (mm)       & 27.07 & 27.07 & 27.07 \\
Lens 2 radius $R_{3}$ (mm)          & -730.28 & -414.39 & -405.13\\
Lens 2 radius $R_{4}$ (mm)          & 273.05 & 239.16 & 206.92 \\
Lens 2 thickness $t_{2}$ (mm)       & 9.00 & 9.00 & 9.00 \\
Lens 2 glass                        & $\mathrm{ADF355}$ & $\mathrm{ADF355}$ & $\mathrm{ADF355}$ \\
Lens 2 Semi-Diameter (mm)                & 49.00 & 49.00 & 49.00 \\
Lens separation $d_{23}$ (mm)       & 13.86 & 13.86 & 13.86 \\
Lens 3 radius $R_{5}$ (mm)          & 174.09 & 195.00 & 174.42 \\
Lens 3 radius $R_{6}$ (mm)          & -611.42 & -381.84 & -395.13 \\
Lens 3 thickness $t_{3}$ (mm)       & 16.00 & 16.00 & 16.00 \\
Lens 3 glass                        &$\mathrm{K\text{-}PFK85}$&$\mathrm{K\text{-}PFK85}$& $\mathrm{K\text{-}PFK85}$\\
Lens 3 Semi-Diameter (mm)            & 45.00 & 45.00 & 45.00 \\
Distance to image plane (mm)        & 83.29 & 80.96 & 79.23 \\
Effective focal length (mm)         & 9127.75 & 9127.2 & 9127.2 \\
Approx $f/\#$                       & 4.25  & 4.25  & 4.25 \\
\midrule
\multicolumn{4}{c}{\textit{Dichroic Elements}} \\
\midrule
Dichroic tilt angle (deg) &  & $45$ & $-45$ \\
Wedge angle (deg)        &  & $8.9\times10^{-2}$ & $-11.7\times10^{-2}$ \\
Aperture (mm)                 &  & $70 \times 97$ & $61 \times 84$ \\
Thickness (mm)            &  & 10.0 & 10.0 \\
Material                  &  & Fused silica & Fused silica \\
Central wavelength ($\mu$m)             &     0.45     &     0.61     &    0.835 \\
Spectral range ($\mu$m)             & 0.35--0.55 & 0.55--0.67 & 0.67--1.00 \\
\bottomrule
\end{tabular}
\end{table*}

    \subsection{Nominal optical performance}\label{subsec:Nominal}

    \begin{figure*}[t]
        \centering           
        \includegraphics[width=0.65\textwidth]{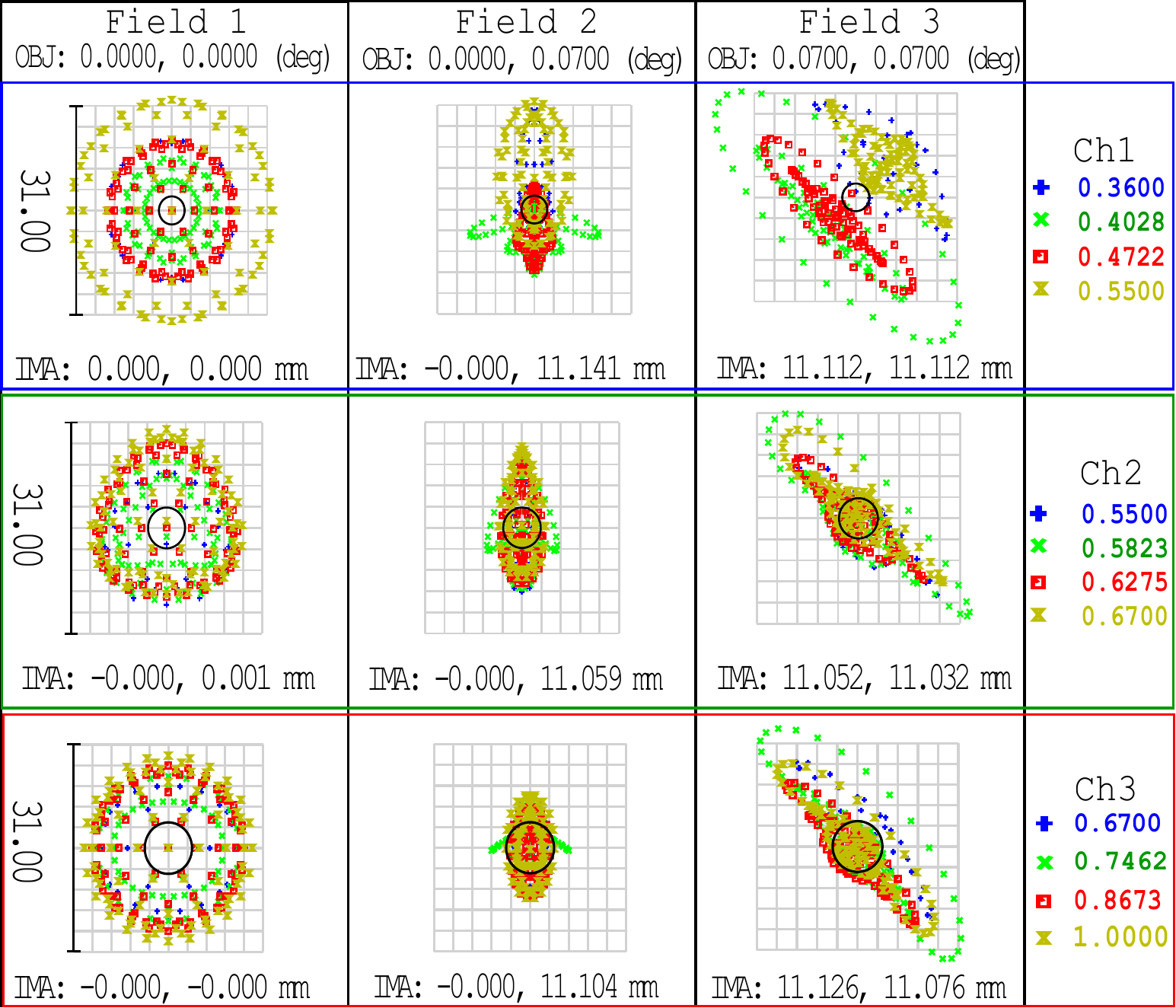}
        \caption{\justifying Spot diagrams from exact ray tracing for the final optical design. The focal reducers were optimized in \texttt{KrakenOS}; ZEMAX\textsuperscript{\textregistered} OpticStudio was used for best focus determination, dichroic wedge optimization, and final fine adjustment. For each spectral channel (rows), three image plane field points are shown: the on-axis field $(0.00,\,0.00)$ and the off-axis fields $(0.00,\,0.07)$ and $(0.07,\,0.07)$. Colors indicate the wavelengths considered in each channel. The scale bar corresponds to $31\,\mu$m, equivalent to $0.7\,\arcsec$, and is included for visual reference only.}
        \label{fig:SPT_all}
    \end{figure*}
            
    The nominal optical performance of the instrument is characterized through the analysis of spot diagrams obtained by exact ray tracing. The evaluation focuses on the size and spatial distribution of the point images, in order to quantify the image quality achieved in each spectral channel. Figure\,\ref{fig:SPT_all} shows the spot diagrams for the three spectral channels of the instrument, arranged by rows: channel\,1 (top), channel\,2 (middle), and channel\,3 (bottom). For each channel, one on-axis field and two off-axis positions are included, allowing a consistent description of the system behavior under both ideal conditions and in regions dominated by off-axis aberrations. All spot diagrams are displayed using a common scale bar of $31\,\mu$m, corresponding to a seeing of $0.7\,\arcsec$, included for visual reference only.
            
    The use of colors allows identification of the different wavelengths considered in each channel, facilitating the assessment of the chromatic performance of the system. For each diagram, the corresponding positions in the object plane and image plane are indicated.
            
    For off-axis fields, the geometric (GEO) spot radius, defined as the maximum distance between the spot-diagram centroid and the outermost ray intersection, may exceed the reference radius associated with a seeing of $0.7\,\arcsec$ ($\sim15.5\,\mu$m). The maximum values obtained are $25.28\,\mu$m in the blue channel, $26.2\,\mu$m in the green channel, and $22.17\,\mu$m in the red channel. This behavior is explained by the presence of isolated marginal rays with low energy contribution, which tend to bias this metric toward larger values. Such an effect is consistent with the expected behavior of multielement systems evaluated off-axis and does not, by itself, imply a significant degradation of energy concentration.

    In contrast, the RMS spot radius remains below $11\,\mu$m for all analyzed fields and channels. This value is comparable to the pixel size of the sCMOS detector ($11\,\mu$m) and ensures that most of the point-image energy is concentrated within an area equivalent to one pixel or its immediate vicinity. As a result, the RMS radius constitutes a more representative metric of the photometric performance and effective image quality of the system. When expressed in angular units using the system plate scale ($22.6\,\arcsec\,\mathrm{mm}^{-1}$), the RMS values correspond to characteristic angular sizes below $\sim0.25\,\arcsec$ in all cases. This sampling corresponds to an oversampled regime with respect to the typical site seeing, indicating that the instrument performance is limited by the atmosphere rather than by the optical design.
            
    Given that the previous analysis based on the RMS radius already establishes that the characteristic angular size of the image is significantly smaller than the representative atmospheric seeing at the site, a complementary metric is introduced to further characterize the energy distribution in the image plane. In particular, the EE50 metric is analyzed, as it is less sensitive to the presence of isolated marginal rays and provides a more robust estimate of the effective image size.
    
\begin{figure*}[t]
        \centering                
        \includegraphics[width=0.7\textwidth]{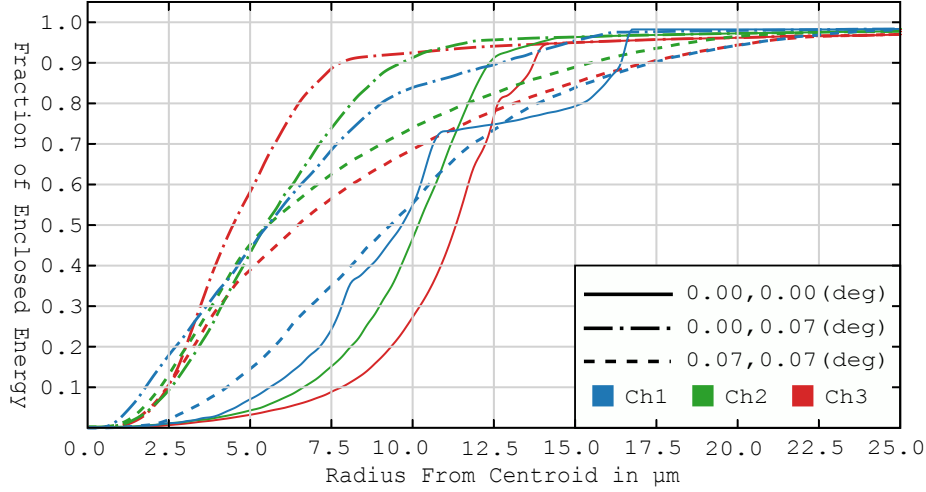}
        \caption{\justifying Encircled geometric energy curves, normalized by the diffraction limit, for the three spectral channels: blue (channel\,1), green (channel\,2), and red (channel\,3). Line styles indicate field positions: solid (on-axis), dash–dotted (off-axis), and dashed (diagonal).}
        \label{fig:EE_All}
    \end{figure*}

    The EE50 calculation is carried out polychromatically. Figure\,\ref{fig:EE_All} displays the encircled geometric energy curves for the three spectral channels, distinguished by representative colors, computed at the same field points considered previously and normalized by the diffraction limit. The horizontal axis represents the cumulative encircled energy fraction, whereas the vertical axis indicates the radius from the spot-diagram centroid, given in units of $\mu$m.
    
    In the encircled-energy analysis, the on-axis field $(0.00,\,0.00)$ is adopted as the reference, since in the nominal design it consistently presents the largest EE50 radii (see Figure\,\ref{fig:EE_All}), thus constituting the most demanding case for the photometric performance of the system. For this field, the obtained EE50 radii are $9.67\,\mu$m for the blue channel, $10.20\,\mu$m for the green channel, and $11.28\,\mu$m for the red channel. These values are comparable to the sCMOS detector pixel size ($11\,\mu$m), indicating that a significant fraction of the point-image energy is concentrated within an area of the order of one pixel.
            
    For off-axis fields, EE50 values are generally equal to or smaller than those corresponding to the central field. For the field $(0.00,\,0.07)$, EE50 radii of $5.56\,\mu$m, $5.48\,\mu$m, and $4.48\,\mu$m are obtained for the blue, green, and red channels, respectively, while for the diagonal field $(0.07,\,0.07)$ the values are $9.31\,\mu$m, $5.56\,\mu$m, and $6.41\,\mu$m. This behavior is consistent with the off-axis energy concentration inferred from the geometric analysis of the spot diagrams.
                    
    When expressed in angular units using the system plate scale ($22.6\,\arcsec\,\mathrm{mm}^{-1}$), the EE50 radii correspond to characteristic angular sizes below $\sim0.26\,\arcsec$ for all channels and fields considered. These values lie well below the typical site seeing, confirming that the instrument performance is dominated by atmospheric effects rather than by limitations of the optical design.
            
    Table\,\ref{tab:NominalMetrics} summarizes the metrics characterizing the nominal performance of the focal reduction system. In the following subsection, the effect of the dichroics on image quality is evaluated, considering wedge optimization as part of the control of off-axis aberrations in a converging beam.
    
    \begin{table}[ht]
        \centering
        \caption{\justifying Nominal optical performance metrics for the three spectral channels and the representative analyzed fields. RMS denotes the root-mean-square spot radius computed from the ray intersections at the image plane, expressed in micrometers.}
        \label{tab:NominalMetrics}
        \begin{tabular}{c c c c}
        \toprule
        \textbf{Channel} & \textbf{Field} & \textbf{RMS [$\mu$m]} & \textbf{EE50 [$\mu$m]} \\
        \midrule
              & $(0.00,\,0.00)$ & 10.51 & 9.67 \\
        Blue  & $(0.00,\,0.07)$ & 6.97  & 5.56 \\
              & $(0.07,\,0.07)$ & 10.81 & 9.31 \\
        \midrule
               & $(0.00,\,0.00)$ & 9.85  & 10.20 \\
        Green  & $(0.00,\,0.07)$ & 5.87  & 5.48 \\
               & $(0.07,\,0.07)$ & 8.06  & 5.56 \\
        \midrule
              & $(0.00,\,0.00)$ & 10.85 & 11.28 \\
        Red   & $(0.00,\,0.07)$ & 4.50  & 4.48 \\
              & $(0.07,\,0.07)$ & 8.75  & 6.41 \\
        \bottomrule
        \end{tabular}
    \end{table}

    \subsection{Effect of dichroic wedging}\label{subsec:Dicacu}

    As described in Section\,\ref{subsec:Dichroics}, the insertion of dichroics in a converging beam breaks the rotational symmetry of the optical system and introduces off-axis aberrations, primarily coma and astigmatism. In the adopted design, these effects are mitigated by incorporating a wedge on the second surface of each dichroic as part of the final numerical refinement.
    
    The wedge angles are optimized using an extended set of representative field points to properly sample the off-axis behavior associated with the effective rectangular geometry of the dichroics. In addition to the three nominal fields, symmetric field points such as $(0.07,\,0.00)$, $(0.00,\,-0.07)$, and $(0.07,\,-0.07)$, as well as intermediate diagonal positions, are included in the optimization.
    
    The final wedge angles are $8.9\times10^{-2}$\,deg for the first dichroic ($\mathrm{D}_1$) and $-11.7\times10^{-2}$\,deg for the second dichroic ($\mathrm{D}_2$). These values restore effective optical-axis coherence in each channel and reduce the off-axis aberrations introduced by transmission through the converging beam. As a consequence, the image quality metrics reported in Section\,\ref{subsec:Nominal} satisfy the established RMS and EE50 criteria across all analyzed fields. Given the sensitivity of system performance to dichroic geometry, the impact of manufacturing tolerances is evaluated in the following subsection.

    \subsection{Tolerance analysis}\label{subsec:TolAnalysis}

    Geometric and material deviations associated with both lenses and dichroics are considered. The corresponding parameters and variation ranges are summarized in Table\,\ref{tab:TolerancesGlobal} as perturbations relative to the nominal design. Tolerances expressed in fringe units are defined with respect to the test wavelength corresponding to each spectral channel. The impact of these perturbations is quantified through the degradation of the EE50 radius, which is adopted as a representative metric of the system's sensitivity to manufacturing errors.
    
    \begin{table}[ht]
        \centering
        \caption{\justifying Tolerance parameters considered for lenses and dichroics. The indicated ranges correspond to deviations relative to the nominal design.}
        \label{tab:TolerancesGlobal}
        \begin{tabular}{l c c}
        \toprule
        \textbf{Element} & \textbf{Parameter} & \textbf{Range} \\
        \midrule
               & Radius of curvature & $\pm 3$ fringes \\
               & Thickness & $\pm 0.05$ mm \\
        Lenses & Wedge & $\leq 0.01$ mm \\
               & Ast+Sph irregularity & $\leq 0.5$ fringes \\
               & Glass homogeneity & $0.005$ to $0.5\%$ \\[2pt]
        \midrule
                  & Thickness & $\pm 0.05$ mm \\
        Dichroics & Wedge & $\leq 0.01$ mm \\
                  & Glass homogeneity & $0.005$ to $0.5\%$ \\
        \bottomrule
        \end{tabular}
    \end{table}
    
    A worst case approach is adopted, such that acceptance of the system under these conditions guarantees its operational feasibility under less unfavorable combinations of manufacturing errors. Within this scheme, the EE50 values obtained for each channel and field are converted into an equivalent FWHM metric. This allows the contributions associated with lens tolerances, dichroic tolerances, the nominal design, and atmospheric seeing to be combined in quadrature. The resulting total FWHM values are summarized in Table\,\ref{tab:performance_canales_total}.
    
    \begin{table}[ht]
        \centering
        \caption{\justifying Total FWHM in arcseconds per channel, obtained as the quadrature sum of nominal performance, lens tolerances, dichroic tolerances, and atmospheric seeing of $1.54\,\arcsec$.}
        \label{tab:performance_canales_total}
        \begin{tabular}{c c c c}
        \toprule
        \textbf{Field} & \textbf{Blue} & \textbf{Green} & \textbf{Red} \\
        \midrule
        $(0.00,\,0.00)$ & 1.73 & 1.90 & 1.92 \\
        $(0.00,\,0.07)$ & 1.61 & 1.65 & 1.60 \\
        $(0.07,\,0.07)$ & 1.72 & 1.65 & 1.68 \\
        \bottomrule
        \end{tabular}
    \end{table}
            
    For the first channel, performance degradation is dominated by tolerances associated with the lenses, while the contribution from the dichroics is negligible. In the on-axis field, which constitutes the most demanding scenario, the EE50 values induced by lens tolerances reach $10.15\,\mu$m, while in the off-axis fields values of $5.84\,\mu$m for $(0.00,\,0.07)$ and $9.23\,\mu$m for $(0.07,\,0.07)$ are obtained. These results remain bounded and are compatible with the sampling and energy concentration requirements established for the system, also reflecting a lower sensitivity to perturbations in off-axis regions.
            
    In the second and third channels, where the light interacts with one and two dichroics respectively, the contribution of these elements becomes comparable to that of the lenses within the total error budget. For the second channel, lens tolerances lead to EE50 values of $14.30\,\mu$m in the central field and $6.98\,\mu$m and $5.08\,\mu$m in the off-axis fields, while dichroic tolerances introduce additional degradations of $11.68\,\mu$m, $6.11\,\mu$m, and $5.51\,\mu$m for the same fields. Similarly, in the third channel, EE50 values due to lenses of $14.63\,\mu$m, $5.24\,\mu$m, and $6.55\,\mu$m are obtained, together with dichroic contributions of $11.41\,\mu$m, $4.09\,\mu$m, and $6.17\,\mu$m, respectively.
    
    These results indicate that the dichroics constitute fabrication-critical components, particularly in the on-axis field, where their impact is maximal and comparable to that of the lenses; nevertheless, even under this most unfavorable scenario, the additional EE50 degradation remains within acceptable margins. In off-axis fields, the behavior is systematically more favorable, with EE50 values lower than those of the central field, reflecting a reduced sensitivity of the system to manufacturing perturbations in these regions. Overall, the tolerance analysis confirms that the final instrument performance remains dominated by factors external to the optical design, most notably atmospheric conditions, rather than by variations introduced during the manufacturing of the optical elements.
    
\section{Discussion}\label{sec:Discusion}

    The comparison between nominal performance and the system response under manufacturing tolerances shows that the optical design of the \textit{OPTICAM-ARG} instrument satisfies the image quality requirements established in the instrumental specifications, even under the worst case scenario considered in the tolerance analysis.
    
    The stable behavior observed across the three spectral channels reflects the effectiveness of the adopted optimization strategy, which combines explicit control of off-axis aberrations with systematic evaluation of fabrication sensitivity. In particular, the optimized dichroic wedging mitigates the aberrations introduced by operation in a converging beam and preserves effective optical-axis coherence in each channel, contributing to uniform image quality across the field of view.
    
    The results also emphasize the sensitivity of system performance to the geometry of both dichroics and focal-reducer lenses. Because these elements directly influence energy concentration and aberration balance, their tolerances represent a critical component of the overall error budget. Within the adopted tolerance ranges, however, the design remains robust and compliant with the established RMS and EE50 criteria.
    
    From the standpoint of field sampling, the off-axis fields impose the most demanding conditions in terms of aberrations, while the on-axis field is adopted as a reference for the global evaluation of the system. This criterion ensures a coherent optical response across the field of view and among the different spectral channels. In a broader context, the instrument design constitutes a natural extension of previously validated multi-band approaches, adapted to the specific optical and mechanical conditions of the JS telescope. In this sense, the proposed scheme is particularly well-suited for high cadence instrumentation aimed at the study of transient phenomena from the southern hemisphere.
    
    From a methodological perspective, the results confirm the effectiveness of combining a physically grounded design stage with a subsequent geometric refinement, allowing the control of dominant aberrations to be decoupled from the fine tuning of image quality. This approach favors interpretable and reproducible solutions and facilitates the implementation of the optical design using open source software tools such as \texttt{KrakenOS} and \texttt{SciPy}, as well as specialized commercial software, in particular ZEMAX\textsuperscript{\textregistered}, enabling cross validation of the design.
    
    Although overall performance remains conditioned by the assumed seeing regime, the results indicate that the system is limited primarily by atmospheric conditions rather than by intrinsic optical design constraints, and can therefore be implemented without the need for additional corrective optical elements.

\section{Conclusions}\label{sec:Conclusiones}

    In this work, we presented the optical design of the high cadence multi-band astronomical camera \textit{OPTICAM-ARG}, optimized for implementation on the Jorge Sahade Telescope. The system satisfies the image quality requirements defined in the instrumental specifications, maintaining stable performance across the three spectral channels and throughout the field of view, even under conservative manufacturing tolerance scenarios and representative atmospheric conditions at the site.
    
     Physically consistent starting points were constructed from geometrical optics principles, incorporating paraxial modeling and third-order corrections, and were subsequently refined through numerical optimization that combined a physically grounded merit function with geometric minimization of the RMS spot radius. A central element of the configuration is the use of dichroic wedging, which effectively compensates for the off-axis aberrations introduced by operation in a converging beam, ensuring consistent optical performance among the spectral channels.
    
    From a methodological perspective, the results demonstrate that combining physically interpretable modeling with structured numerical refinement constitutes an efficient and reproducible strategy for the development of complex astronomical instrumentation. The complete workflow, spanning initial design construction, analytical modeling, and numerical optimization, was implemented within the open source \texttt{KrakenOS} framework, which integrates optical models and optimization schemes in a modular and reproducible manner. ZEMAX\textsuperscript{\textregistered} OpticStudio was employed in a complementary role for focus adjustment, dichroic wedge optimization, and final geometric refinement, providing independent verification of the design.
    
    The proposed methodology is extensible to more complex optical systems, including additional multi-channel architectures, aspheric implementations, or telescopes with faster focal ratios, and is compatible with nondeterministic optimization approaches for exploring highly nonlinear parameter spaces. Overall, \textit{OPTICAM-ARG} establishes a robust and reproducible framework for high cadence multi-band photometric instrumentation in the southern hemisphere, achieving atmospheric-limited performance without the need for additional corrective optical elements under the adopted design assumptions.


\section{FACILITIES}

Jorge Sahade Telescope (CASLEO; planned), \textit{OPTICAM-ARG} camera


\renewcommand{\refname}{REFERENCES}
\bibliography{opticamarg_refs}

@article{simons1995technical,
  title={Technical note no. 30},
  author={Simons, DA},
  journal={Gemini Observatory},
  year={1995}
}

@article{hoschel2019genetic,
  author    = {Kaspar H{\"o}schel and Vasudevan Lakshminarayanan},
  title     = {Genetic algorithms for lens design: a review},
  journal   = {Journal of Optics},
  volume    = {48},
  number    = {1},
  pages     = {134--144},
  year      = {2019},
  doi       = {10.1007/s12596-018-0497-3},
  url       = {https://doi.org/10.1007/s12596-018-0497-3},
  issn      = {0974-6900}
}

@book{rutten1988telescope,
  title={Telescope Optics: Evaluation and Design},
  author={Rutten, H.G.J. and van Venrooij, M.A.M. and Berry, R. and Lucas, D.},
  isbn={9780943396187},
  lccn={88014243},
  url={https://books.google.com.mx/books?id=x3LvAAAAMAAJ},
  year={1988},
  publisher={Willmann-Bell}
}

@book{wilson2004reflecting,
  title={Reflecting Telescope Optics I: Basic Design Theory and its Historical Delvelopment},
  author={Wilson, Raymond N},
  year={2004},
  publisher={Springer}
}

@book{sasian2019introduction,
  title = {Introduction to Lens Design},
  ISBN = {9781108494328},
  url = {http://dx.doi.org/10.1017/9781108625388},
  DOI = {10.1017/9781108625388},
  publisher = {Cambridge University Press},
  author = {Sasián,  José},
  year = {2019},
  month = sep 
}

@article{dhillon2021hipercam,
  title={HiPERCAM: a quintuple-beam, high-speed optical imager on the 10.4-m Gran Telescopio Canarias},
  author={Dhillon, VS and Bezawada, N and Black, M and Dixon, SD and Gamble, T and Gao, X and Henry, DM and Kerry, P and Littlefair, SP and Lunney, DW and others},
  journal={Monthly Notices of the Royal Astronomical Society},
  volume={507},
  number={1},
  pages={350--366},
  year={2021},
  publisher={Oxford University Press}
}

@inproceedings{butler2012first,
  title={First light with RATIR: an automated 6-band optical/NIR imaging camera},
  author={Butler, Nat and Klein, Chris and Fox, Ori and Lotkin, Gennadiy and Bloom, Josh and Prochaska, J Xavier and Ramirez-Ruiz, Enrico and Jos{\'e}, A and Georgiev, Leonid and Gonz{\'a}lez, Jes{\'u}s and others},
  booktitle={Ground-based and Airborne Instrumentation for Astronomy IV},
  volume={8446},
  pages={336--342},
  year={2012},
  organization={SPIE}
}

@misc{miller2025lasillaschmidtsouthern,
      title={The La Silla Schmidt Southern Survey}, 
      author={Adam A. Miller and Natasha S. Abrams and Greg Aldering and Shreya Anand and Charlotte R. Angus and Iair Arcavi and Charles Baltay and Franz E. Bauer and Daniel Brethauer and Joshua S. Bloom and Hemanth Bommireddy and Marcio Catelan and Ryan Chornock and Peter Clark and Thomas E. Collett and Georgios Dimitriadis and Sara Faris and Francisco Forster and Anna Franckowiak and Christopher Frohmaier and Lluıs Galbany and Renato B. Galleguillos and Ariel Goobar and Claudia P. Gutierrez and Saarah Hall and Erica Hammerstein and Kenneth R. Herner and Isobel M. Hook and Macy J. Huston and Joel Johansson and Charles D. Kilpatrick and Alex G. Kim and Robert A. Knop and Marek P. Kowalski and Lindsey A. Kwok and Natalie LeBaron and Kenneth W. Lin and Chang Liu and Jessica R. Lu and Wenbin Lu and Ragnhild Lunnan and Kate Maguire and Lydia Makrygianni and Raffaella Margutti and Dan Maoz and Patrik Milan Veres and Thomas Moore and A. J. Nayana and Matt Nicholl and Jakob Nordin and Giuliano Pignata and Abigail Polin and Dovi Poznanski and Jose L. Prieto and David L. Rabinowitz and Nabeel Rehemtulla and Mickael Rigault and Dan Ryczanowski and Nikhil Sarin and Steve Schulze and Ved G. Shah and Xinyue Sheng and Samuel P. R. Shilling and Brooke D. Simmons and Avinash Singh and Graham P. Smith and Mathew Smith and Jesper Sollerman and Maayane T. Soumagnac and Christopher W. Stubbs and Mark Sullivan and Aswin Suresh and Benny Trakhtenbrot and Charlotte Ward and Eli Wiston and Helen Xiong and Yuhan Yao and Peter E. Nugent},
      year={2025},
      eprint={2503.14579},
      archivePrefix={arXiv},
      primaryClass={astro-ph.IM},
      url={https://arxiv.org/abs/2503.14579}, 
}

@article{sullivan2019pyvista,
   doi = {10.21105/joss.01450},
  url = {https://doi.org/10.21105/joss.01450},
  year = {2019},
  month = {May},
  publisher = {The Open Journal},
  volume = {4},
  number = {37},
  pages = {1450},
  author = {Bane Sullivan and Alexander Kaszynski},
  title = {{PyVista}: {3D} plotting and mesh analysis through a streamlined interface for the {Visualization Toolkit} ({VTK})},
  journal = {Journal of Open Source Software}
}

@book{vtkBook,
 author    = "Will Schroeder and Ken Martin and Bill Lorensen",
  title     = "The Visualization Toolkit (4th ed.)",
  publisher = "Kitware",
  year      = "2006",
  isbn      = "978-1-930934-19-1",
}

@misc{OpenAI_ChatGPT,
  title = {ChatGPT},
  author = {OpenAI},
  year = {2024},
  url = {https://openai.com/chatgpt}
}

@article{hunter2007matplotlib,
  author={Hunter, John D.},
  journal={Computing in Science \& Engineering}, 
  title={Matplotlib: A 2D Graphics Environment}, 
  year={2007},
  volume={9},
  number={3},
  pages={90-95},
  doi={10.1109/MCSE.2007.55}}

@article{harris2020array,
 title = {Array programming with {NumPy}},
 author = {Charles R. Harris and K. Jarrod Millman and St{\'{e}}fan J. van der Walt and Ralf Gommers and Pauli Virtanen and David Cournapeau and Eric Wieser and Julian Taylor and Sebastian Berg and Nathaniel J. Smith and Robert Kern and Matti Picus and Stephan Hoyer and Marten H. van Kerkwijk and Matthew Brett and Allan Haldane and Jaime Fern{\'{a}}ndez del R{\'{i}}o and Mark Wiebe and Pearu Peterson and Pierre G{\'{e}}rard-Marchant and Kevin Sheppard and Tyler Reddy and Warren Weckesser and Hameer Abbasi and Christoph Gohlke and Travis E. Oliphant},
 year  = {2020},
 month = {sep},
 journal = {Nature},
 volume  = {585},
 number = {7825},
 pages  = {357--362},
 doi = {10.1038/s41586-020-2649-2},
 publisher = {Springer Science and Business Media {LLC}},
 url = {https://doi.org/10.1038/s41586-020-2649-2}
}

@ARTICLE{Shahbaz+15,
       author = {{Shahbaz}, T. and {Linares}, M. and {Nevado}, S.~P. and {Rodr{\'\i}guez-Gil}, P. and {Casares}, J. and {Dhillon}, V.~S. and {Marsh}, T.~R. and {Littlefair}, S. and {Leckngam}, A. and {Poshyachinda}, S.},
        title = "{The binary millisecond pulsar PSR J1023+0038 during its accretion state - I. Optical variability}",
      journal = {\mnras},
         year = 2015,
        month = nov,
       volume = {453},
       number = {4},
        pages = {3461-3473},
          doi = {10.1093/mnras/stv1686},
archivePrefix = {arXiv},
       eprint = {1507.07473},
 primaryClass = {astro-ph.HE},
       adsurl = {https://ui.adsabs.harvard.edu/abs/2015MNRAS.453.3461S}
}

@ARTICLE{Veresvarska+25,
       author = {{Veresvarska}, M. and {Scaringi}, S. and {Littlefield}, C. and {de Martino}, D. and {Knigge}, C. and {Paice}, J. and {Altamirano}, D. and {Castro}, A. and {Michel}, R. and {Castro Segura}, N. and {Echevarr{\'\i}a}, J. and {Groot}, P.~J. and {Hern{\'a}ndez Santisteban}, J.~V. and {Irving}, Z.~A. and {Altamirano-D{\'e}vora}, L. and {Sahu}, A. and {Buckley}, D.~A.~H. and {Vincentelli}, F.},
        title = "{DW Cnc: a micronova with a negative superhump and a flickering spin}",
      journal = {\mnras},
         year = 2025,
        month = may,
       volume = {539},
       number = {3},
        pages = {2424-2434},
          doi = {10.1093/mnras/staf412},
archivePrefix = {arXiv},
       eprint = {2503.07704},
 primaryClass = {astro-ph.HE},
       adsurl = {https://ui.adsabs.harvard.edu/abs/2025MNRAS.539.2424V}
}

@ARTICLE{Castro-Segura+2025,
       author = {{Castro Segura}, N. and {Irving}, Z.~A. and {Vincentelli}, F.~M. and {Altamirano}, D. and {Tampo}, Y. and {Knigge}, C. and {Pelisoli}, I. and {Coppejans}, D.~L. and {Rawat}, N. and {Castro}, A. and {Sahu}, A. and {Hern{\'a}ndez Santisteban}, J.~V. and {Kimura}, M. and {Veresvarska}, M. and {Michel}, R. and {Scaringi}, S. and {Najera}, M.~R.},
        title = "{Bridging the gap: OPTICAM reveals the hidden spin of the WZ Sge star GOTO 065054.49+593624.51}",
      journal = {\mnras},
         year = 2025,
        month = jul,
       volume = {541},
       number = {1},
        pages = {L28-L34},
          doi = {10.1093/mnrasl/slaf038},
archivePrefix = {arXiv},
       eprint = {2501.11669},
 primaryClass = {astro-ph.HE},
       adsurl = {https://ui.adsabs.harvard.edu/abs/2025MNRAS.541L..28C}
}

@ARTICLE{Irving+25,
       author = {{Irving}, Z.~A. and {Castro Segura}, N. and {Altamirano}, D. and {Scaringi}, S. and {Veresvarska}, M. and {Vincentelli}, F. and {de Martino}, D. and {Buckley}, D.~A.~H. and {Castro}, A. and {Michel}, R.},
        title = "{OPTICAM reveals hints of cyclotron emission from the intermediate polar V709 Cas}",
      journal = {\mnras},
         year = 2026,
        month = jan,
       volume = {545},
       number = {3},
          eid = {staf2177},
        pages = {staf2177},
          doi = {10.1093/mnras/staf2177},
       adsurl = {https://ui.adsabs.harvard.edu/abs/2026MNRAS.545f2177I}
}

@article{castro-segura-2025,
    author = {Castro Segura, N and Irving, Z A and Vincentelli, F M and Altamirano, D and Tampo, Y and Knigge, C and Pelisoli, I and Coppejans, D L and Rawat, N and Castro, A and Sahu, A and Santisteban, J V Hernández and Kimura, M and Veresvarska, M and Michel, R and Scaringi, S and Najera, M R},
    title = {Bridging the gap: OPTICAM reveals the hidden spin of the WZ Sge star GOTO 065054.49+593624.51},
    journal = {Monthly Notices of the Royal Astronomical Society: Letters},
    volume = {541},
    number = {1},
    pages = {L28-L34},
    year = {2025},
    month = {04},
    issn = {1745-3925},
    doi = {10.1093/mnrasl/slaf038},
    url = {https://doi.org/10.1093/mnrasl/slaf038},
    eprint = {https://academic.oup.com/mnrasl/article-pdf/541/1/L28/63002903/slaf038.pdf},
}

@ARTICLE{1996AJ....111.1748F,
       author = {{Fukugita}, M. and {Ichikawa}, T. and {Gunn}, J.~E. and {Doi}, M. and {Shimasaku}, K. and {Schneider}, D.~P.},
        title = "{The Sloan Digital Sky Survey Photometric System}",
      journal = {\aj},
         year = 1996,
        month = apr,
       volume = {111},
        pages = {1748},
          doi = {10.1086/117915},
       adsurl = {https://ui.adsabs.harvard.edu/abs/1996AJ....111.1748F}
}

@ARTICLE{2007MNRAS.378..825D,
       author = {{Dhillon}, V.~S. and {Marsh}, T.~R. and {Stevenson}, M.~J. and {Atkinson}, D.~C. and {Kerry}, P. and {Peacocke}, P.~T. and {Vick}, A.~J.~A. and {Beard}, S.~M. and {Ives}, D.~J. and {Lunney}, D.~W. and {McLay}, S.~A. and {Tierney}, C.~J. and {Kelly}, J. and {Littlefair}, S.~P. and {Nicholson}, R. and {Pashley}, R. and {Harlaftis}, E.~T. and {O'Brien}, K.},
        title = "{ULTRACAM: an ultrafast, triple-beam CCD camera for high-speed astrophysics}",
      journal = {\mnras},
         year = 2007,
        month = jul,
       volume = {378},
       number = {3},
        pages = {825-840},
          doi = {10.1111/j.1365-2966.2007.11881.x},
archivePrefix = {arXiv},
       eprint = {0704.2557},
 primaryClass = {astro-ph},
       adsurl = {https://ui.adsabs.harvard.edu/abs/2007MNRAS.378..825D}
}

@manual{andor_catalogo,
  title        = {Marana sCMOS
Ultimate Sensitivity Back-illuminated
sCMOS for Astronomy \& Physical Sciences},
  author       = {Oxford Instruments},
  year         = {2021},
  url          = {https://pdf.directindustry.es/pdf-en/andor-technology/marana-scmos/34591-1084955.html}
}

@article{Howard:85,
author = {James W. Howard},
journal = {Appl. Opt.},
number = {23},
pages = {4265--4268},
publisher = {Optica Publishing Group},
title = {Formulas for the coma and astigmatism of wedge prisms used in converging light},
volume = {24},
month = {Dec},
year = {1985},
url = {https://opg.optica.org/ao/abstract.cfm?URI=ao-24-23-4265},
doi = {10.1364/AO.24.004265},
}

@book{conrady1957applied,
  author = {Conrady, Alexander Eugen},
  year = {1992},
  title = {Applied Optics and Optical Design, Part One},
  publisher = {Douver Publications, Inc}
}

@book{conrady1960applied,
  title = {Applied Optics and Optical Design, Part Two},
  author = {Conrady, Alexander Eugen and Kingslake, Rudolf},
  year = {1992},
  publisher = {Douver Publications, Inc}
}

@book{kingslake1978lens,
  title = {Lens Design Fundamentals},
  author = {Kingslake, Rudolf},
  year = {1978},
  publisher = {Academic press}
}

@book{bentley2012field,
  title = {Field Guide to Lens Design},
  ISBN = {9780819491657},
  url = {http://dx.doi.org/10.1117/3.934997},
  DOI = {10.1117/3.934997},
  publisher = {SPIE},
  author = {Bentley,  Julie and Olson,  Craig},
  year = {2012},
  month = dec 
}

@article{castro+19,
  title={Opticam: a triple-camera optical system designed to explore the fastest timescales in astronomy},
  author={Castro, A and Altamirano, D and Michel, R and Gandhi, P and Hern{\'a}ndez Santisteban, JV and Echevarr{\'\i}a, J and Tejada, C and Knigge, C and Sierra, G and Colorado, E and others},
  journal={Revista mexicana de astronom{\'\i}a y astrof{\'\i}sica},
  volume={55},
  number={2},
  pages={363--376},
  year={2019},
  publisher={Instituto de Astronom{\'\i}a}
}

@article{mammanamediciones,
  title={MEDICIONES DE SEEING EN EL COMPLEJO ASTRON{\'O}MICO EL LEONCITO MEDIANTE LA T{\'E}CNICA DIMM, Y SU CORRELACI{\'O}N CON PAR{\'A}METROS METEOROL{\'O}GICOS},
  author={Mammana, Luis},
  journal = {Technical Report CASLEO},
  year={2018}
}

@article{10.1119/1.1970159,
    author = {Halbach, Klaus},
    title = {Matrix Representation of Gaussian Optics},
    journal = {American Journal of Physics},
    volume = {32},
    number = {2},
    pages = {90-108},
    year = {1964},
    month = {02},
    issn = {0002-9505},
    doi = {10.1119/1.1970159},
    url = {https://doi.org/10.1119/1.1970159},
    eprint = {https://pubs.aip.org/aapt/ajp/article-pdf/32/2/90/11884950/90\_1\_online.pdf},
}

@book{welford2017aberrations,
  title={Aberrations of optical systems},
  author={Welford, Walter Thompson},
  year={2017},
  publisher={Routledge}
}

@book{kidger2001fundamental,
  title={Fundamental optical design},
  author={Kidger, Michael J},
  year={2001},
  publisher={SPIE Optical Engineering Press}
}

@ARTICLE{2020SciPy-NMeth,
  author  = {Virtanen, Pauli and Gommers, Ralf and Oliphant, Travis E. and
            Haberland, Matt and Reddy, Tyler and Cournapeau, David and
            Burovski, Evgeni and Peterson, Pearu and Weckesser, Warren and
            Bright, Jonathan and {van der Walt}, St{\'e}fan J. and
            Brett, Matthew and Wilson, Joshua and Millman, K. Jarrod and
            Mayorov, Nikolay and Nelson, Andrew R. J. and Jones, Eric and
            Kern, Robert and Larson, Eric and Carey, C J and
            Polat, {\.I}lhan and Feng, Yu and Moore, Eric W. and
            {VanderPlas}, Jake and Laxalde, Denis and Perktold, Josef and
            Cimrman, Robert and Henriksen, Ian and Quintero, E. A. and
            Harris, Charles R. and Archibald, Anne M. and
            Ribeiro, Ant{\^o}nio H. and Pedregosa, Fabian and
            {van Mulbregt}, Paul and {SciPy 1.0 Contributors}},
  title   = {{{SciPy} 1.0: Fundamental Algorithms for Scientific
            Computing in Python}},
  journal = {Nature Methods},
  year    = {2020},
  volume  = {17},
  pages   = {261--272},
  adsurl  = {https://rdcu.be/b08Wh},
  doi     = {10.1038/s41592-019-0686-2},
}

@article{herrera2022krakenos,
  title={KrakenOS: Python-based general exact ray tracing library},
  author={Herrera, Joel and Guerrero, Carlos A and Najera, Morgan R and Sotelo-Burke, Anais and Plauchu-Frayn, Ilse},
  journal={Optical Engineering},
  volume={61},
  number={1},
  pages={015101--015101},
  year={2022},
  publisher={Society of Photo-Optical Instrumentation Engineers}
}

@book{malacara2003handbook,
  title={Handbook of optical design},
  author={Malacara, Zacarias and Malacara-Hern{\'a}ndez, Daniel and Malacara-Hern{\'a}ndez, Zacar{\'\i}as},
  year={2003},
  publisher={CRC press}
}

\section{ACKNOWLEDGEMENTS}

We thank the referees for their constructive comments and suggestions that helped improve the clarity and quality of this manuscript. We also thank the editors and the editorial assistant for their support and efficient handling of the review process. This collaboration is supported by a tripartite Letter of Intent between the University of Southampton, the Instituto de Astronomía of UNAM, and the Complejo Astronómico El Leoncito (CASLEO), signed by all partners in December 2023, with a renewed version currently in progress. We gratefully acknowledge these institutions for providing the institutional framework for the collaborative development of the \textit{OPTICAM-ARG} project. We acknowledge the technical staff of CASLEO for their support during the development of this work. In particular, we thank Roberto Francisco Sánchez and Miguel Ángel Giménez Ferrando for providing the technical measurements of the Jorge Sahade Telescope, which were essential for the optical design presented here. We acknowledge the collaboration of Gerardo Sierra and Javier Hernández, who are currently developing the mechanical design associated with the optical system presented in this article. We also thank Sergio Cellone for providing the seeing data used as reference in the performance analysis of the instrument. We thanks Noel Castro Segura and Federico Vincentelli for very useful discussions. MRN and EL acknowledge support from the DGAPA–PAPIIT grant IT120125. MRN also acknowledges SECIHTI for financial support through a doctoral scholarship (CVU: 1044690). A.C acknowledges support from the Royal Society Newton International Fellowships NF170803, AL\textbackslash221034, and AL\textbackslash24100056. D.A. thanks STFC’s Early Technology Development Capital Funding Call 2022 for their support buying the essential equipment to build OPTICAM-ARG.
\section*{Software}
The software used in this work includes \texttt{KrakenOS}\,\citep{herrera2022krakenos}, \texttt{NumPy}\,\citep{harris2020array}, \texttt{Matplotlib}\,\citep{hunter2007matplotlib}, \texttt{SciPy}\,\citep{2020SciPy-NMeth}, \texttt{PyVista}\,\citep{sullivan2019pyvista}, \texttt{VTK}\,\citep{vtkBook}, and \texttt{ZEMAX OpticStudio}\textsuperscript{\textregistered}. Additionally, language editing assistance was supported using ChatGPT\,\citep{OpenAI_ChatGPT}.









\end{document}